\documentstyle[epsfig]{aipproc}

\begin{document}

\def\gr{\hbox{ \raisebox{-1.0mm}{$\stackrel{>}{\sim}$} }}
\def\kr{\hbox{ \raisebox{-1.0mm}{$\stackrel{<}{\sim}$} }}

\title{Near-infrared polarimetric observations of the afterglow of GRB 000301C}

\author{B. Stecklum$^a$, O. Fischer$^b$, S. Klose$^a$,
        R. Mundt$^c$, and \mbox{C. Bailer-Jones$^{c,}$
        \thanks{Visiting Astronomer, German-Spanish Astronomical
        Centre, Calar Alto,  operated by the Max-Planck-Institute for
        Astronomy, Heidelberg, jointly with the Spanish National Commission
        for Astronomy}}}
\address{$^a$Th\"uringer Landessternwarte, Tautenburg, Germany \\
         $^b$Universit\"atssternwarte, Jena, Germany \\
         $^c$Max-Planck-Institut f\"ur Astronomie, Heidelberg, Germany}

\maketitle

\begin{abstract}
Based on near-infrared polarimetric observations  we constrain the
degree of linear polarization of the afterglow light of GRB 000301C 
to less than 30\% \ 1.8 days after the burst.
\end{abstract}

                   \section{Introduction}

The question whether a GRB is always accompanied by a collimated
outflow, often called a jet, is one of the key issues of current GRB
research.  Because of relativistic effects a collimated explosion will
reduce the deduced energy release of a GRB by the beaming
factor. Theoretical considerations have demonstrated that in the case
of a collimated outflow the afterglow light can be partly linearly
polarized and the degree of linear polarization $p$ should vary
with time /1-4/.

In 1999 we established a Target of Opportunity (ToO) program aiming at
the measurement of the linear polarization of GRB afterglows. The
project is being carried out at the 3.5-m telescope at Calar Alto,
Spain, utilizing the near-infrared camera Omega~Cass /5/.  This
instrument is equipped with a $1024\,\times\,1024$ HAWAII array. Thus
far, it was employed in wide field mode ($0\farcs3$/pixel), yielding a
field of view of $\sim5\arcmin\,\times\,5\arcmin$. The observations
were performed in the $K'$ band, using wire-grid polarizers to obtain
images at  four position angles (0, 45, 90, and 135\arcdeg). The
limiting magnitude of the combined images amounts to $K'\sim19$.

The burst GRB 000301C was the first burst for which we can 
constrain the degree of linear polarization of the afterglow light. 

                   \section{The burst GRB 000301C}

The GRB 000310C was detected with \it RXTE, Ulysses, \rm and \it NEAR
\rm  on March 1,  2000, at 9:51 UTC /6/. In the high-energy band 
it lasted $\sim$2 seconds  /7/. The optical afterglow was soon
detected on images taken on 2000 March 3.2 UT at $R = 20.3 \pm
0.5$ /8/. The redshift of the host galaxy turned out to be  2.04
/7/. A break in the afterglow lightcurve was seen some days after the 
burst suggesting that this burst was accompanied by a jet
/9, 10/. Of particular interest is the possible detection of a
microlensing event in the afterglow lightcurve /11/.

\begin{figure}[t!]
\centerline{\epsfig{file=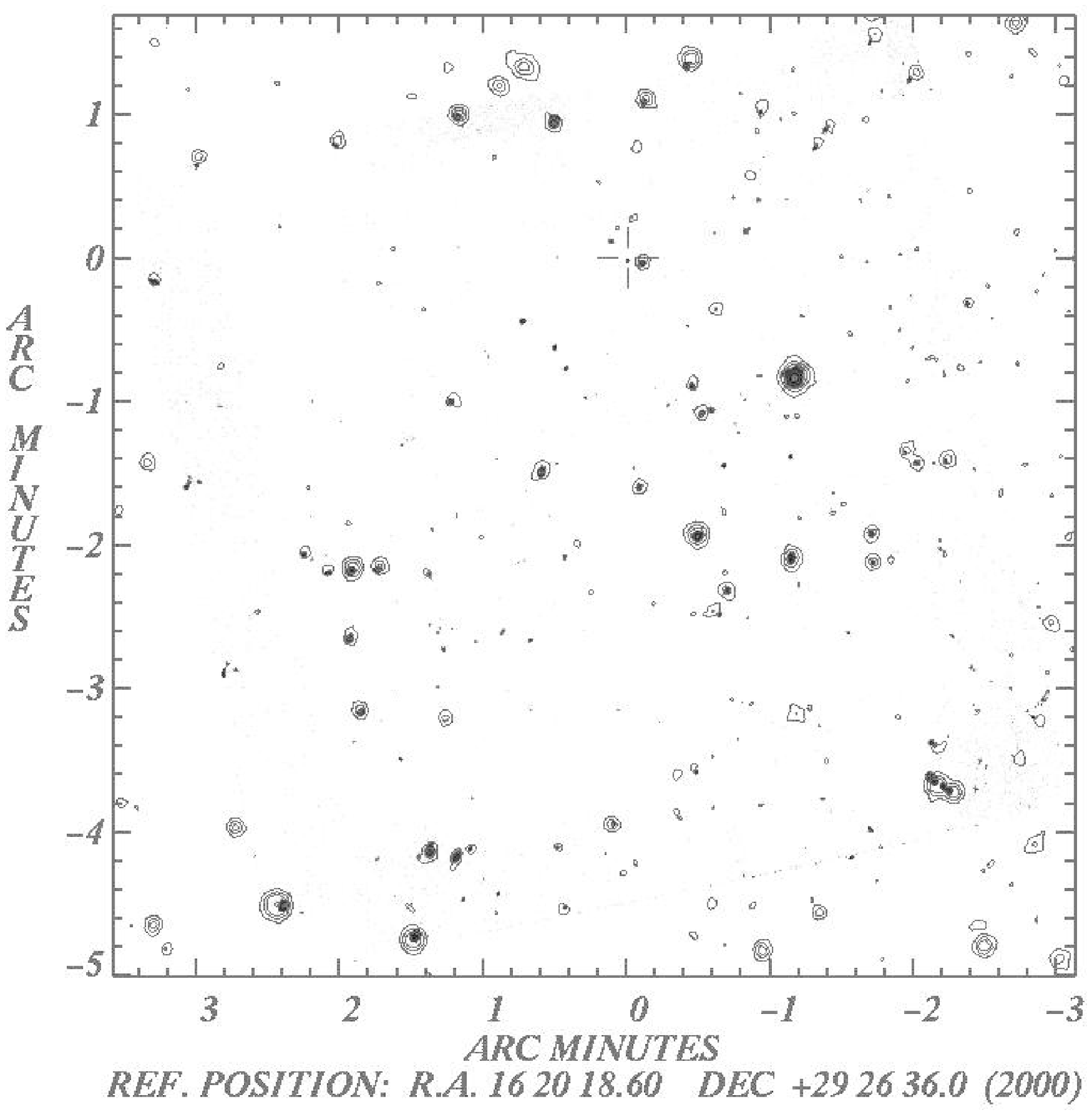,height=3.5in,width=3.5in}}
\vspace{10pt}
\caption{The afterglow of GRB 000301C was imaged 1.8 days after the
burst with the Calar Alto 3.5-m telescope at a magnitude of $K'$=17.5
/12/.  Displayed here is the combined $K'$-band image after  adding all
frames taken at four different polarization angles. The GRB afterglow
is indicated by a cross. Contour lines represent the 
overplotted DSS-2 red image of the field.}
\end{figure}

The  polarimetric data were acquired at Calar Alto on March 3, 5:00 UT
when the  GRB afterglow was already at a magnitude of $K'$=17.5
(Fig. 1). The instrumental and interstellar contribution to the linear
polarization were corrected by assuming a zero net polarization of all
stars in the field. Based on aperture photometry of all well-isolated
stars we can constrain $p$ of the GRB afterglow to be less than 30\%
(Fig.\,2). This result is in agreement with predictions  according to
which $p$ should never exceed about 20\% /1, 2/.  We note, however,
that our result might still be influenced by residual instrumental
polarization which is currently under investigation.

\begin{figure}[t!]
\centerline{\epsfig{file=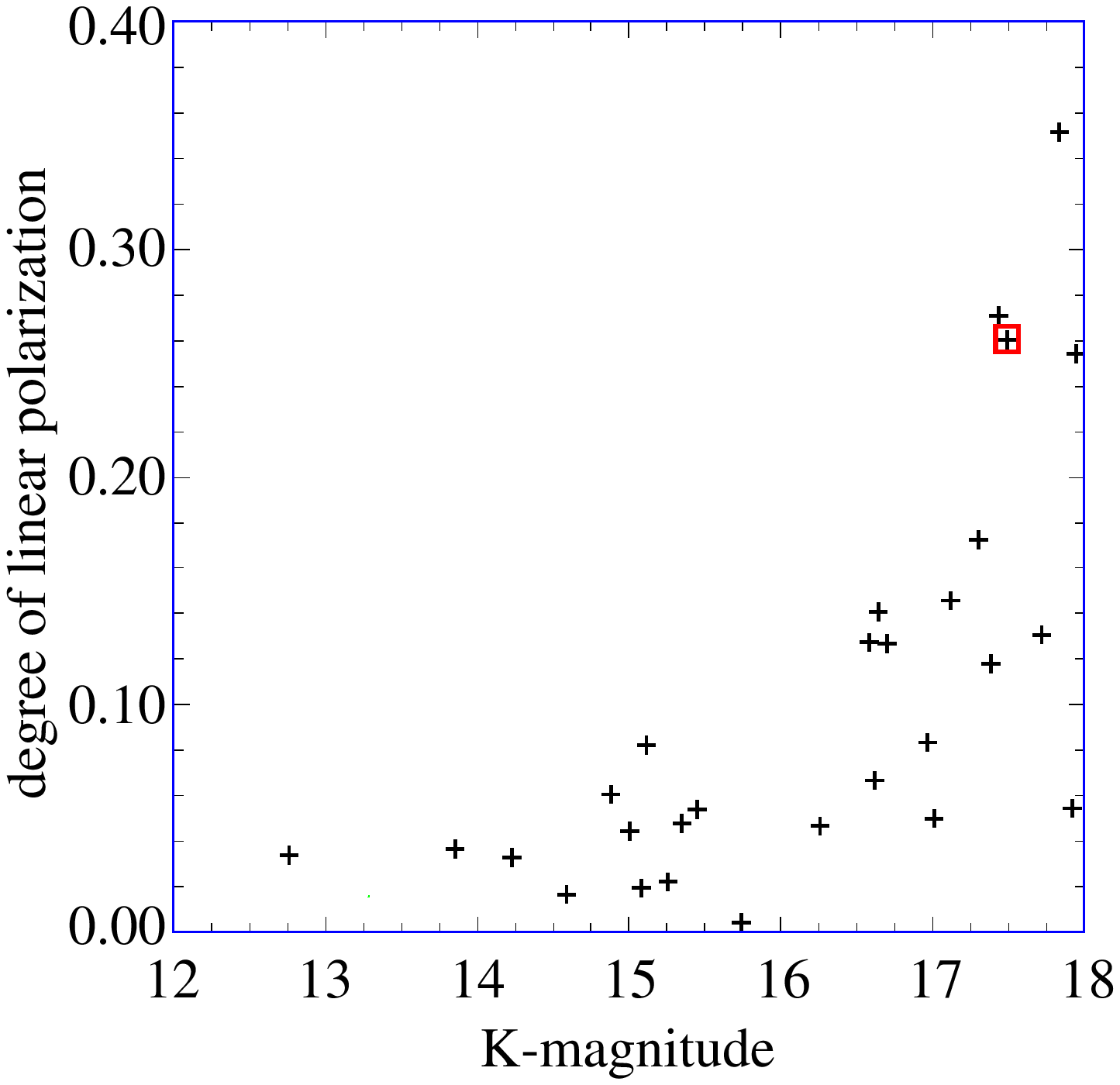,height=3.0in,width=3.0in}}
\caption{Polarization $vs.$ magnitude diagram for all stars 
in the field ($p=1$ corresponds to 100\% polarization).
The GRB afterglow is indicated by a square.}
\end{figure}

                   \section{Future work}

The project entered its second year of operation in January 2001. Our
results show that a GRB afterglow has to be brighter than  $K'\sim16$
in order to measure, or to constrain, its degree of linear
polarization with an error of $\lesssim10\%$ using Omega~Cass. From
\it HETE \rm 2 we expect rapid alerts and accurate GRB locations which
would allow to use the high-resolution mode of Omega~Cass. The
instrumental capabilities for this mode and the larger brightness of
the GRB afterglows due to the shorter response time suggest that our
goal of measuring $p$ and its temporal variation will be achievable in
the near future.


\end{document}